\documentclass{article}
\usepackage{spconf,amsmath,graphicx,booktabs}
\usepackage[hidelinks]{hyperref}

\title{NormToken: SPEAKER- AND DURATION-NORMALIZED SEMANTIC SPEECH TOKENS VIA ITERATIVE S2U--T2U REFINEMENT}
\name{Hanlin Zhang$^{1,*,\ddagger}$, Daxin Tan$^{2,*}$,
Dehua Tao$^{2,*}$, Chengxi Deng$^{3}$, Xiao Chen$^{2,\dagger}$,
Linqi Song$^{1,\dagger}$
\thanks{$^{*}$Equal contribution.\quad
$^{\dagger}$Corresponding authors.\quad
$^{\ddagger}$This work was done during an internship at Huawei.}}
\address{$^{1}$Department of Computer Science, City University of Hong Kong\\
$^{2}$AI Lab, Leibniz Research Center, Huawei\\
$^{3}$Chinese University of Hong Kong\\
\texttt{\{hanlzhang8-c@my., linqi.song@\}cityu.edu.hk}\\
\texttt{\{chen.xiao2, tan.daxin1\}@huawei.com}}
\begin{document}
\ninept
\maketitle
\begin{abstract}
Semantic speech tokens should preserve linguistic content while suppressing
utterance-specific acoustic and duration variation. However, existing
speech-to-unit (S2U) tokenizers often retain speaker-related acoustic
characteristics and duration information. To address this issue, we propose
NormToken, an iterative semantic token purification framework that alternates
S2U and text-to-unit (T2U) training. In each iteration, the T2U model produces text-derived tokens to supervise a newly initialized S2U tokenizer. The resulting S2U tokens are then used as targets for the next T2U iteration.
This cycle drives the two models toward a shared, text-predictable token space.
Experiments on Mandarin and English demonstrate improved S2U--T2U agreement and
parallel-utterance token consistency. De-tokenizers trained on the initial and refined 
tokens further show that refined tokens maintain comparable WER and CER while improving
speaker similarity in both voice cloning and text-to-speech synthesis. In voice
cloning, refined tokens produce speaking rates closer to the acoustic reference,
suggesting reduced dependence on source duration information.
Audio samples are available at \url{https://hanlin1004.github.io/normtoken_demopage/} .
\end{abstract}
%

%\vspace{-0.5em}

\section{Introduction}
\label{sec:intro}

Semantic speech tokens provide a compact interface for speech recognition,
translation, and generation~\cite{hsu2021hubert,lee2022direct}. Ideally, they
should represent what is said rather than how or by whom it is spoken.
However, existing speech-to-unit (S2U) tokenizers often retain non-linguistic
properties of the input speech. Consequently, utterances with the
same linguistic content may produce different token sequences when 
spoken by different speakers or with different speaking rates. Such variations are undesirable for semantic
tokens, which should primarily preserve linguistic content.
ContentVec shares the motivation of suppressing speaker-dependent variation,
but operates on continuous self-supervised representations~\cite{qian2022contentvec}. 
In contrast, we focus on low-rate discrete tokens
that can be modeled directly for speech generation. Our key observation is
that linguistic content is predictable from text, whereas speaker
characteristics and speaking rate cannot be determined from text alone.
We therefore use agreement with a text-to-unit (T2U) model as a text-based normalization
signal to purify the semantic tokens generated by S2U sequences 
while suppressing speaker- and duration-dependent variation.

Based on this observation, we propose NormToken, an iterative semantic token
purification framework. Starting from an initial S2U tokenizer \(S_0\), NormToken first trains a T2U model \(T_0\) to predict its deduplicated token sequences. The resulting text-derived tokens are then used to train a newly initialized S2U tokenizer \(S_1\) with connectionist temporal classification (CTC) loss~\cite{graves2006ctc}. This process is iteratively repeated, where each refined S2U tokenizer provides new targets for the next T2U model, forming the cycle
$S_0\!\rightarrow T_0\!\rightarrow S_1\!\rightarrow T_1\!\rightarrow\cdots$.
Instead of merely generating tokens, the T2U model provides text-predictable supervision that progressively refines the S2U tokenizer.

Our contributions are threefold:
(i) we characterize cross-speaker sequence inconsistency and utterance-duration dependence in discrete semantic tokens;
(ii) we propose NormToken, an iterative S2U--T2U procedure that uses text-predictable supervision to refine the speech tokenizer; and
(iii) we demonstrate across Mandarin and English that the proposed refinement improves cross-model agreement and cross-speaker consistency, enhances speaker similarity in text-to-speech (TTS) and voice conversion (VC), and preserves intelligibility.

\begin{figure*}[t]
    \centering
    \includegraphics[
        width=0.90\textwidth,
        height=0.35\textheight
    ]{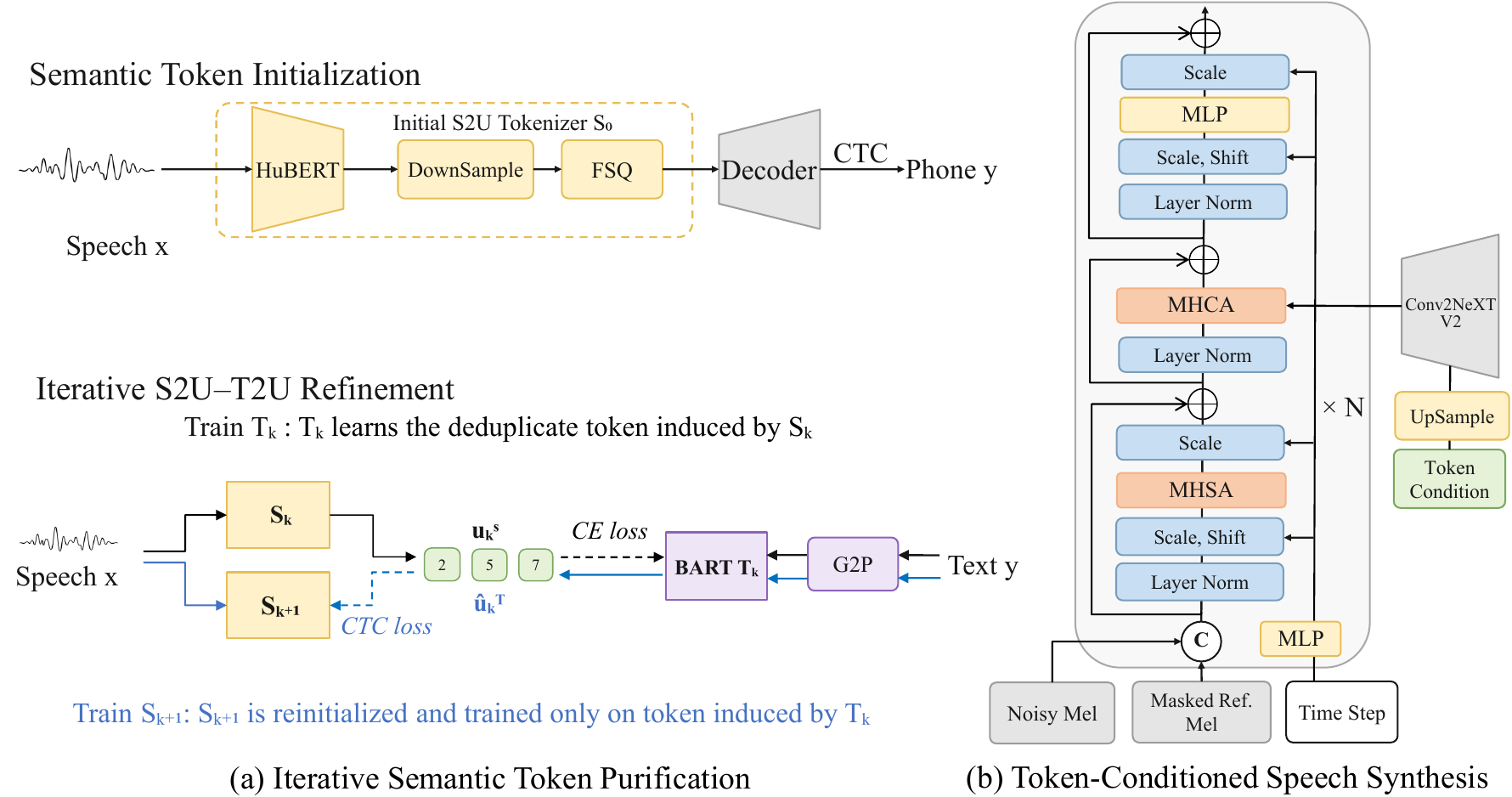}
    \caption{Overview of the proposed framework. (a) NormToken first trains an
    initial S2U tokenizer and then performs iterative S2U--T2U refinement. At
    each iteration, the T2U model is trained on deduplicated S2U tokens, and the
    T2U-generated tokens are used to train the reinitialized S2U tokenizer for the
    next iteration. (b) The token-conditioned de-tokenizer synthesizes speech from
    either S2U or T2U tokens.}
    \label{fig:method}
\end{figure*}

\vspace{-0.5em}

\section{Related Work}
\label{sec:related}

\subsection{Invariant and efficient semantic speech tokenization}

Semantic speech tokens are commonly obtained by quantizing self-supervised
representations such as HuBERT~\cite{hsu2021hubert}. To improve speaker and
noise invariance, ContentVec disentangles speaker information from continuous
representations, R-Spin enforces consistency across perturbed views, and PINT
leverages parallel utterances and augmentations~\cite{qian2022contentvec,
chang2024rspin,wagner2026pint}. Duration variation is another source of token
inconsistency. TASTE uses text-aligned segment aggregation, while FlexiCodec
learns adaptive frame rates for more efficient tokenization
~\cite{tseng2025taste,li2026flexicodec}. NormToken addresses both sources of variation through iterative semantic token purification, combining deduplication for duration normalization with T2U supervision for text-guided refinement. It requires neither speaker
labels nor parallel utterances with the same content, as text predictability
itself serves as the normalization criterion.

\subsection{Text-to-unit Modeling}
T2U models have been used to synthesize training data from text. BT4ST
generates source speech through target-to-unit and unit-to-speech models,
whereas DUB directly uses predicted source units to augment a unit-to-text
translation model~\cite{fang2023bt4st,zhang2023dub}. Zeng et al. convert sampled
text spans into speech-token spans to construct large-scale synthetic
interleaved data for speech-text pre-training~\cite{zeng2025synthetic}. These methods keep the S2U tokenizer fixed and use T2U predictions to train downstream models. In contrast, NormToken treats T2U-generated tokens as text-derived supervision to iteratively refine the S2U tokenizer itself.

%\vspace{-0.5em}

\section{NormToken}
\label{sec:method}

\subsection{Overview and Initialization}

Figure~\ref{fig:method} summarizes the full framework. Given paired speech and
transcript $(\mathbf{x},\mathbf{y})$, NormToken adopts an iterative refinement strategy, 
which trains an S2U tokenizer $S_k$ and a T2U model $T_k$ alternatively. 
Here, $k$ denotes the refinement iteration. Unlike joint optimization, $S_k$ is obtained before $T_k$ is trained, and $T_k$ then provides supervision for $S_{k+1}$.

The initial tokenizer $S_0$ follows the semantic branch of
DSA-Tokenizer~\cite{tao2024tone}. A HuBERT encoder~\cite{hsu2021hubert}
extracts contextual speech representations, which are temporally downsampled
and quantized using finite scalar quantization (FSQ)~\cite{mentzer2024fsq}.
FSQ uses 4 scalar dimensions with 8 levels per dimension, yielding
$8^4=4{,}096$ discrete units. The tokenizer operates at 25~Hz. During initial
training, an auxiliary phoneme CTC decoder encourages the quantized
representations to preserve linguistic content. This decoder is discarded
after $S_0$ is trained; only the discrete FSQ indices are used to initialize
the alternating refinement procedure.

\subsection{Iterative S2U--T2U Refinement}

At iteration $k$, the deduplication operator $\mathcal{D}$ removes consecutive
repetitions from the current S2U sequence $S_k(\mathbf{x})$, producing a
shorter target sequence while preserving unit order. A BART encoder--decoder~\cite{lewis2020bart} 
takes the grapheme-to-phoneme (G2P)-converted phoneme
sequence as input and predicts the deduplicated S2U token sequence:

\begin{equation}
    \mathbf{u}_k^S = \mathcal{D}(S_k(\mathbf{x})), \qquad
    \mathcal{L}_{T}^k
    = -\log p_{T_k}\!\left(
        \mathbf{u}_k^S \mid \operatorname{G2P}(\mathbf{y})
      \right).
\end{equation}

Thus, $T_k$ first learns the token sequence generated by the current
$S_k$. At inference, it autoregressively decodes a text-derived token sequence
$\hat{\mathbf{u}}_k^T$ from the transcript. Because this prediction is
conditioned on text rather than the source waveform, it does not have access to 
the source-speaker characteristics or utterance-specific timing information.

We then train a newly initialized S2U tokenizer $S_{k+1}$ rather than fine-tuning $S_k$. The new S2U
model applies a CTC classifier directly to the downsampled HuBERT
representations and uses the same 4,096-unit vocabulary. 
It is trained only on the text-derived token sequence generated by $T_k$:

\begin{equation}
    \hat{\mathbf{u}}_k^T = T_k(\operatorname{G2P}(\mathbf{y})), \qquad
    \mathcal{L}_{S}^{k+1}
    = -\log p_{S_{k+1}}^{\mathrm{CTC}}\!\left(
        \hat{\mathbf{u}}_k^T \mid \mathbf{x}
      \right).
\end{equation}

CTC provides monotonic alignment without frame-level correspondence. 
Reinitialization is important because $S_{k+1}$ must relearn the
speech-to-unit mapping from the text-constrained targets instead of directly
inheriting the previous S2U model. Once training finishes, the output of
$S_{k+1}$ is decoded and deduplicated to train $T_{k+1}$, and the same two-step
procedure repeats. Since the supervision transferred from T2U to S2U is conditioned solely on text, the iterative cycle encourages token patterns that are recoverable from linguistic content while suppressing text-unpredictable speaker and duration cues.

The two refinement steps play complementary roles. Deduplication reduces
repetitions caused by duration variation, whereas T2U supervision further
suppresses token variations that cannot be predicted from the transcript. The
alternating procedure can therefore be viewed as gradually aligning the S2U
and T2U token spaces toward a shared, text-predictable representation.

\subsection{Token-Conditioned Speech Synthesis}

To test generation utility, we train separate token-to-speech de-tokenizers with the same
architecture for iteration 0 and the final iteration $K$. This avoids unfairly penalizing refined tokens due to a decoder that is optimized only for the initial token distribution. Each de-tokenizer is trained on a mixture of S2U
sequences extracted from speech and T2U sequences predicted from text at the
corresponding iteration.

We adopt the CA-F5-TTS de-tokenizer used in
Speech-Omni-Lite~\cite{tao2026speechomnilite,chen2025f5tts}. The discrete unit
sequence is upsampled and supplied as the semantic condition, while a masked
reference mel spectrogram provides acoustic context. A conditional flow
matching objective reconstructs the target mel spectrogram. At evaluation,
S2U tokens extracted from a source utterance are used for VC, whereas T2U tokens
predicted from text are used for TTS. The shared de-tokenizer design therefore
tests both speech-derived and text-predicted units under the same synthesis
backbone.

%\vspace{-0.5em}

\section{Experimental Setup}
\label{sec:experiments}

\textbf{Training data.}
We sample approximately 8,000 hours of paired speech and text. The English
portion draws from LibriSpeech~\cite{panayotov2015librispeech},
GigaSpeech~\cite{chen2021gigaspeech},
Libri-Heavy~\cite{kang2024libriheavy}, Common
Voice~\cite{ardila2020commonvoice}, and The People's
Speech~\cite{galvez2021peoplesspeech}. The Mandarin portion draws from
AISHELL-2~\cite{du2018aishell2}, WenetSpeech~\cite{zhang2022wenetspeech}, and
MagicData-RAMC~\cite{yang2022magicdata}. The same multilingual pool is used
for both S2U and T2U training. For T2U, each transcript is converted by G2P
into a phoneme sequence as input, while the deduplicated unit sequence
produced by the corresponding S2U tokenizer serves as the target. Each refined
S2U tokenizer follows the iterative refinement procedure described in
Section~\ref{sec:method}. The initial and refined de-tokenizers are trained
separately with the same architecture and the same approximately 100,000 hours
of Mandarin and English speech from Emilia~\cite{he2024emilia}. Each
de-tokenizer uses a mixture of S2U- and T2U-derived sequences from its
corresponding token space.

\textbf{Baselines.}
For speech synthesis, we compare with MaskGCT~\cite{wang2024maskgct},
CosyVoice 2~\cite{du2024cosyvoice2}, and
CosyVoice 3~\cite{du2025cosyvoice3}. We also include DualCodec with
MaskGCT~\cite{li2025dualcodec}, TaDiCodec-AR~\cite{wang2025tadicodec},
Spark-TTS~\cite{wang2025sparktts}, and X-VC~\cite{zheng2026xvc}.
For token consistency, we include
StableToken~\cite{song2025stabletoken}, SAC~\cite{chen2025sac},
FlexiCodec~\cite{li2026flexicodec}, R-Spin~\cite{chang2024rspin}, and the
semantic tokenizer of CosyVoice 3. Table~\ref{tab:generation} reports all TTS
baseline results and the X-VC result from their original papers; the remaining
VC results are evaluated using our implementation.

\textbf{Evaluation metrics.}
We evaluate cross-model agreement using token-sequence WER and BLEU, and
evaluate synthesis quality on the Seed-TTS-Eval benchmark~\cite{anastassiou2024seedtts}
using English WER, Chinese CER, and speaker similarity (SIM). We further
evaluate token consistency on VCTK~\cite{veaux2017vctk} using UED
\cite{gat2022augmentation} and SelfBLEU-4 over token sequences extracted from
utterances sharing the same linguistic content. For UED, we compute pairwise
edit distances between token sequences and normalize each distance by the
reference sequence length. SelfBLEU-4 treats each token sequence as the
hypothesis and the remaining token sequences from the same-content group as
references, computes sentence BLEU with uniform 1--4-gram weights and
method-1 smoothing, and averages the resulting scores across token sequences.
Speaking rate is computed for each utterance as its phone count divided by its
duration in seconds.

\section{Results and Discussion}

\begin{figure*}[t]
    \centering
    \includegraphics[width=0.90\textwidth]{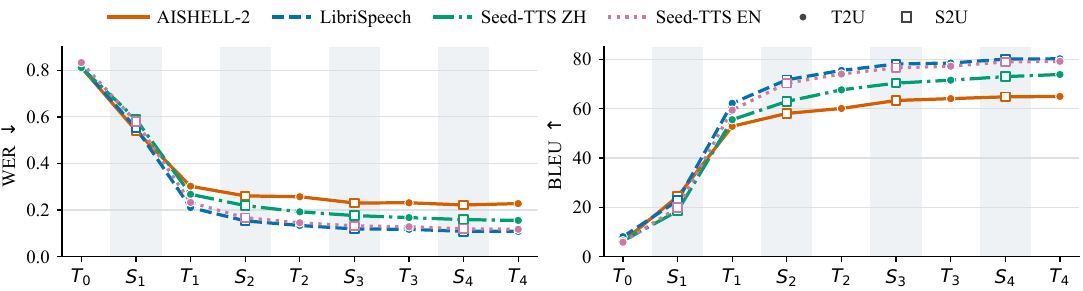}
    \caption{S2U--T2U agreement measured by WER and BLEU over iterative refinement on four test sets.}
    \label{fig:alignment}
    %\vspace{-0.5em}
\end{figure*}

\subsection{Iterative Cross-Model Alignment}

At each iteration, we measure S2U--T2U agreement by comparing T2U outputs
with S2U-derived targets and S2U outputs with T2U-derived targets.
Figure~\ref{fig:alignment} shows that most alignment gains occur within the
first two refinement rounds, after which both metrics approach a plateau.
Based on this observation, we set $K=4$ for all experiments. The largest improvement follows the first T2U update, suggesting that
text-conditioned prediction effectively filters token variations that are not
well aligned with linguistic content. The subsequent improvement of the
newly initialized $S_2$ trained with $T_1$-generated targets shows that the
improved text alignment can be propagated back to the S2U tokenizer. More
generally, the improvements from each $T_k$ are largely preserved when the
corresponding T2U-generated targets are used to train $S_{k+1}$, indicating
that refinement improves the S2U tokenizer itself rather than merely improving
the T2U predictor. The similar trajectories
across all four test sets further demonstrate consistent behavior across
languages and corpus conditions.
From $T_0$ to $T_4$, WER decreases by 72.0--86.7\% relative to $T_0$, while BLEU
increases by 58.88--73.39 points. By iteration 4, S2U and T2U differ by at most
0.0058 WER and 0.85 BLEU, indicating a highly aligned token space across
languages and domains. 

%\vspace{-0.5em}

\subsection{Parallel-Utterance Token Consistency}

We group VCTK utterances with the same linguistic content to measure whether
semantic tokens remain consistent across different recordings. Relative to the
initial tokenizer, the refined tokenizer reduces UED by 82.7\% and raises
SelfBLEU-4 by 67.13 points. Its SelfBLEU-4 score is substantially higher than
those of the listed baselines, showing that iterative refinement produces more
consistent token sequences across same-content utterances.

UED captures global sequence edits, whereas SelfBLEU-4 measures local token
overlap. Improving both metrics indicates that refinement aligns global
sequence structure and local token patterns. These consistency improvements do
not come at the cost of linguistic information, as NormToken maintains
comparable WER and CER in Table~\ref{tab:generation}. The joint improvements in
consistency and generation indicate that refinement reduces utterance-specific
variation in semantic tokens while preserving linguistic information.

\begin{table}[t]
    \caption{Token consistency across parallel VCTK utterances. SelfBLEU-4 is
    reported on a 0--100 scale.}
    \label{tab:parallel-consistency}
    \centering
    \small
    \setlength{\tabcolsep}{6pt}
    \begin{tabular}{lrr}
        \toprule
        Tokenizer & UED $\downarrow$ & SelfBLEU-4 $\uparrow$ \\
        \midrule
        StableToken & 344.54 & 30.95 \\
        SAC semantic & 409.72 & 4.37 \\
        CosyVoice 3 & 404.70 & 10.52 \\
        FlexiCodec (12.5 Hz) & 428.93 & 2.37 \\
        R-Spin (2048) & \underline{241.42} & \underline{76.64} \\
        ContentVec & 320.96 & 40.11 \\
        \midrule
        Ours (Initial) & 344.61 & 27.04 \\
        Ours (Refined) & \textbf{59.44} & \textbf{94.17} \\
        \bottomrule
    \end{tabular}
\end{table}

%\vspace{-0.5em}

\subsection{Generation Utility of Refined Tokens}

\begin{table}[t]
    \caption{Multi-reference VC and TTS results on Seed-TTS-Eval. Best results are
    bolded; second-best results are underlined.}
    \label{tab:generation}
    \centering
    \small
    \renewcommand{\arraystretch}{0.92}
    \begin{tabular*}{\columnwidth}{@{\extracolsep{\fill}}lrrrr@{}}
        \toprule
        & \multicolumn{2}{c}{English} & \multicolumn{2}{c}{Chinese} \\
        \cmidrule(lr){2-3}
        \cmidrule(lr){4-5}
        Model
        & WER $\downarrow$ & SIM $\uparrow$
        & CER $\downarrow$ & SIM $\uparrow$ \\
        \midrule
        \multicolumn{5}{@{}l}{\textbf{\textit{Text-to-Speech}}} \\
        MaskGCT & 2.62 & \underline{0.71} & 2.27 & \underline{0.77} \\
        CosyVoice2 & 2.57 & 0.65 & 1.45 & 0.75 \\
        CosyVoice3 & 2.02 & \textbf{0.72} & \underline{1.16} & \textbf{0.78} \\
        DualCodec-MaskGCT & 4.18 & 0.68 & 1.85 & 0.75 \\
        TaDiCodec-AR-0.5B & 3.88 & 0.65 & \textbf{1.15} & 0.75 \\
        Spark-TTS & 1.98 & 0.58 & 1.20 & 0.67 \\
        Ours (Initial, T2U) & \textbf{1.80} & 0.63 & 1.52 & 0.72 \\
        Ours (Refined, T2U) & \underline{1.97} & \underline{0.71} & 1.78 & \textbf{0.78} \\
        \midrule
        \multicolumn{5}{@{}l}{\textbf{\textit{Voice Cloning}}} \\
        CosyVoice2 & 3.48 & 0.52 & 2.91 & 0.72  \\
        CosyVoice3 & \textbf{2.74} & 0.56 & \textbf{2.21} & \underline{0.75} \\
        X-VC & 3.14 & 0.62 & \underline{2.65} & 0.72 \\
        Ours (Initial, S2U) & 3.00 & \underline{0.63} & \underline{2.65} & 0.73 \\
        Ours (Refined, S2U) & \underline{2.94} & \textbf{0.71} & 2.76 & \textbf{0.77} \\
        \bottomrule
    \end{tabular*}
\end{table}

Refined tokens are evaluated in conditional speech synthesis, where semantic
tokens are combined with acoustic reference information. In TTS, text-derived
T2U tokens provide linguistic content, while in VC, source-speech S2U tokens
provide linguistic content. In both cases, the reference speech provides
acoustic context, allowing us to evaluate whether refinement improves the
utility of semantic tokens for generation.

Table~\ref{tab:generation} shows that refinement improves speaker similarity
while preserving intelligibility. For TTS, refinement slightly increases
English WER from 1.80 to 1.97, while improving speaker similarity from 0.63 to
0.71. It also achieves the highest Chinese speaker similarity (0.78), tied with
CosyVoice3. For VC, refined S2U tokens improve speaker similarity from 0.63 to
0.71 in English and from 0.73 to 0.77 in Chinese, while maintaining comparable
WER and CER.

Since both de-tokenizers use the same architecture and training data before and
after refinement, the improved SIM suggests that the gain comes from improved
token quality rather than decoder adaptation. These results indicate that
refinement produces more suitable semantic tokens for generation, improving
speaker similarity while preserving linguistic content.

\subsection{Reference-Conditioned Speaking Rate Analysis}

\begin{figure}[t]
    \centering
    \includegraphics[width=\columnwidth]{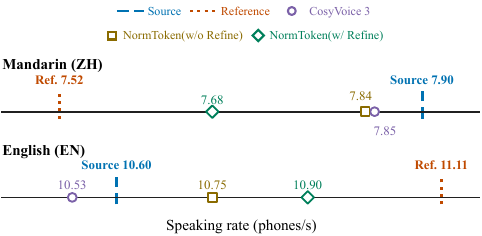}
    \caption{Speaking rates of VC outputs relative to the source utterance and
acoustic reference.}
    \label{fig:rate-transfer}
\end{figure}

In VC, the source utterance provides linguistic content, while the acoustic
reference provides the target-speaker condition. Figure~\ref{fig:rate-transfer}
shows that the speaking rates of CosyVoice 3 and NormToken without refinement
remain closer to the source in both languages. After refinement, this pattern
reverses: the speaking rate produced by NormToken moves closer to the acoustic
reference than to the source utterance. This reversal suggests that refined
tokens retain less source-side duration information, making the generated
speech less dependent on the source speaking rate and supporting the intended
duration-normalization effect. With this effect, our zero-shot TTS and VC models
better transfer prosodic characteristics from the reference speech rather than
merely reproducing its timbre.
Corresponding audio samples are available on our
\href{https://hanlin1004.github.io/normtoken_demopage/}{demopage}.

\section{Conclusion}

NormToken addresses inconsistency in discrete semantic tokens through
iterative S2U--T2U alignment, progressively driving both models toward a
shared, text-predictable token space. Across Mandarin and English, refinement
improves cross-model agreement and parallel-utterance consistency while
reducing speaker-related variation and duration dependence. Importantly,
NormToken improves speaker similarity for both TTS and VC in both languages
while maintaining comparable WER and CER. These results show that NormToken
preserves linguistic content while reducing utterance-specific information,
leading to more effective semantic tokens for reference-conditioned speech
generation.

\bibliographystyle{IEEEbib}
\bibliography{strings,refs}

@article{hsu2021hubert,
      title={HuBERT: Self-Supervised Speech Representation Learning by Masked Prediction of Hidden Units}, 
      author={Wei-Ning Hsu and others},
      year={2021},
      journal={arXiv preprint arXiv:2106.07447},
}

@article{tseng2025taste,
      title={TASTE: Text-Aligned Speech Tokenization and Embedding for Spoken Language Modeling}, 
      author={Liang-Hsuan Tseng and others},
      year={2026},
      journal={arXiv preprint arXiv:2504.07053},
}

@article{li2026flexicodec,
      title={FlexiCodec: A Dynamic Neural Audio Codec for Low Frame Rates}, 
      author={Jiaqi Li and others},
      year={2026},
      journal={arXiv preprint arXiv:2510.00981},
}

@article{chang2024rspin,
      title={R-Spin: Efficient Speaker and Noise-invariant Representation Learning with Acoustic Pieces}, 
      author={Heng-Jui Chang and others},
      year={2024},
      journal={arXiv preprint arXiv:2311.09117},
}

@inproceedings{qian2022contentvec,
      title={ContentVec: An Improved Self-Supervised Speech Representation by Disentangling Speakers},
      author={Kaizhi Qian and others},
      booktitle={Proceedings of the 39th International Conference on Machine Learning},
      pages={18003--18017},
      year={2022},
      publisher={PMLR},
}

@article{wagner2026pint,
      title={Content is What Remains: Invariant Speech Tokenization from Parallel Utterances}, 
      author={Laurin Wagner and others},
      year={2026},
      journal={arXiv preprint arXiv:2607.19033},
}

@inproceedings{fang2023bt4st,
    title = "Back Translation for Speech-to-text Translation Without Transcripts",
    author = "Fang, Qingkai and others",
    editor = "Rogers, Anna  and
      Boyd-Graber, Jordan  and
      Okazaki, Naoaki",
    booktitle = "Proceedings of the 61st Annual Meeting of the Association for Computational Linguistics (Volume 1: Long Papers)",
    month = jul,
    year = "2023",
    address = "Toronto, Canada",
    publisher = "Association for Computational Linguistics",
    url = "https://aclanthology.org/2023.acl-long.251/",
    doi = "10.18653/v1/2023.acl-long.251",
    pages = "4567--4587"
}

@article{zhang2023dub,
      title={DUB: Discrete Unit Back-translation for Speech Translation}, 
      author={Dong Zhang and others},
      year={2023},
      journal={arXiv preprint arXiv:2305.11411},
}

@inproceedings{zeng2025synthetic,
      title={Scaling Speech-Text Pre-training with Synthetic Interleaved Data},
      author={Aohan Zeng and others},
      booktitle={The Thirteenth International Conference on Learning Representations},
      year={2025},
      url={https://proceedings.iclr.cc/paper_files/paper/2025/hash/7b5ae891000049b91b3b62de596b1560-Abstract-Conference.html},
}

@article{lee2022direct,
      title={Direct speech-to-speech translation with discrete units}, 
      author={Ann Lee and others},
      year={2022},
      journal={arXiv preprint arXiv:2107.05604},
}

@inproceedings{graves2006ctc,
  author = {Graves, Alex and others},
  title = {Connectionist temporal classification: labelling unsegmented sequence data with recurrent neural networks},
  year = {2006},
  isbn = {1595933832},
  publisher = {Association for Computing Machinery},
  address = {New York, NY, USA},
  url = {https://doi.org/10.1145/1143844.1143891},
  doi = {10.1145/1143844.1143891},
  booktitle = {Proceedings of the 23rd International Conference on Machine Learning},
  pages = {369--376},
  numpages = {8},
  location = {Pittsburgh, Pennsylvania, USA},
  series = {ICML '06}
}

@article{lewis2020bart,
      title={BART: Denoising Sequence-to-Sequence Pre-training for Natural Language Generation, Translation, and Comprehension}, 
      author={Mike Lewis and others},
      year={2019},
      journal={arXiv preprint arXiv:1910.13461},
}

@article{mentzer2024fsq,
      title={Finite Scalar Quantization: VQ-VAE Made Simple}, 
      author={Fabian Mentzer and others},
      year={2023},
      journal={arXiv preprint arXiv:2309.15505},
}

@article{chen2025f5tts,
      title={F5-TTS: A Fairytaler that Fakes Fluent and Faithful Speech with Flow Matching}, 
      author={Yushen Chen and others},
      year={2025},
      journal={arXiv preprint arXiv:2410.06885},
}

@article{tao2026speechomnilite,
      title={Speech-Omni-Lite: Portable Speech Interfaces for Vision-Language Models}, 
      author={Dehua Tao and others},
      year={2026},
      journal={arXiv preprint arXiv:2603.09627},
}

@inproceedings{panayotov2015librispeech,
  author={Panayotov, Vassil and others},
  booktitle={2015 IEEE International Conference on Acoustics, Speech and Signal Processing (ICASSP)}, 
  title={Librispeech: An ASR corpus based on public domain audio books}, 
  year={2015},
  volume={},
  number={},
  pages={5206-5210},
  doi={10.1109/ICASSP.2015.7178964}}

@inproceedings{chen2021gigaspeech,
   title={GigaSpeech: An Evolving, Multi-Domain ASR Corpus with 10,000 Hours of Transcribed Audio},
   url={http://dx.doi.org/10.21437/Interspeech.2021-1965},
   DOI={10.21437/interspeech.2021-1965},
   booktitle={Interspeech 2021},
   publisher={ISCA},
   author={Chen, Guoguo and others},
   year={2021},
   month=Aug,
   pages={3670--3674}
}

@article{kang2024libriheavy,
      title={Libriheavy: a 50,000 hours ASR corpus with punctuation casing and context}, 
      author={Wei Kang and others},
      year={2024},
      journal={arXiv preprint arXiv:2309.08105},
}

@article{ardila2020commonvoice,
      title={Common Voice: A Massively-Multilingual Speech Corpus}, 
      author={Rosana Ardila and others},
      year={2020},
      journal={arXiv preprint arXiv:1912.06670},
}

@article{du2018aishell2,
      title={AISHELL-2: Transforming Mandarin ASR Research Into Industrial Scale}, 
      author={Jiayu Du and others},
      year={2018},
      journal={arXiv preprint arXiv:1808.10583},
}

@article{zhang2022wenetspeech,
      title={WenetSpeech: A 10000+ Hours Multi-domain Mandarin Corpus for Speech Recognition}, 
      author={Binbin Zhang and others},
      year={2022},
      journal={arXiv preprint arXiv:2110.03370},
}

@article{yang2022magicdata,
      title={Open Source MagicData-RAMC: A Rich Annotated Mandarin Conversational(RAMC) Speech Dataset}, 
      author={Zehui Yang and others},
      year={2022},
      journal={arXiv preprint arXiv:2203.16844},
}

@article{galvez2021peoplesspeech,
      title={The People's Speech: A Large-Scale Diverse English Speech Recognition Dataset for Commercial Usage}, 
      author={Daniel Galvez and others},
      year={2021},
      journal={arXiv preprint arXiv:2111.09344},
}

@article{anastassiou2024seedtts,
      title={Seed-TTS: A Family of High-Quality Versatile Speech Generation Models}, 
      author={Philip Anastassiou and others},
      year={2024},
      journal={arXiv preprint arXiv:2406.02430},
}

@misc{veaux2017vctk,
  author = {Veaux, Christophe and others},
  title = {{CSTR VCTK} Corpus: English Multi-Speaker Corpus for {CSTR} Voice
           Cloning Toolkit},
  year = {2017}
}

@article{gat2022augmentation,
      title={Augmentation Invariant Discrete Representation for Generative Spoken Language Modeling}, 
      author={Itai Gat and others},
      year={2023},
      journal={arXiv preprint arXiv:2209.15483},
}

@article{he2024emilia,
      title={Emilia: An Extensive, Multilingual, and Diverse Speech Dataset for Large-Scale Speech Generation}, 
      author={Haorui He and others},
      year={2024},
      journal={arXiv preprint arXiv:2407.05361},
}

@article{wang2024maskgct,
      title={MaskGCT: Zero-Shot Text-to-Speech with Masked Generative Codec Transformer}, 
      author={Yuancheng Wang and others},
      year={2024},
      journal={arXiv preprint arXiv:2409.00750},
}

@article{du2024cosyvoice2,
      title={CosyVoice 2: Scalable Streaming Speech Synthesis with Large Language Models}, 
      author={Zhihao Du and others},
      year={2024},
      journal={arXiv preprint arXiv:2412.10117},
}

@article{du2025cosyvoice3,
      title={CosyVoice 3: Towards In-the-wild Speech Generation via Scaling-up and Post-training}, 
      author={Zhihao Du and others},
      year={2025},
      journal={arXiv preprint arXiv:2505.17589},
}

@article{li2025dualcodec,
      title={DualCodec: A Low-Frame-Rate, Semantically-Enhanced Neural Audio Codec for Speech Generation}, 
      author={Jiaqi Li and others},
      year={2025},
      journal={arXiv preprint arXiv:2505.13000},
}

@article{wang2025tadicodec,
      title={TaDiCodec: Text-aware Diffusion Speech Tokenizer for Speech Language Modeling}, 
      author={Yuancheng Wang and others},
      year={2025},
      journal={arXiv preprint arXiv:2508.16790},
}

@article{wang2025sparktts,
      title={Spark-TTS: An Efficient LLM-Based Text-to-Speech Model with Single-Stream Decoupled Speech Tokens}, 
      author={Xinsheng Wang and others},
      year={2025},
      journal={arXiv preprint arXiv:2503.01710},
}

@article{zheng2026xvc,
      title={X-VC: Zero-shot Streaming Voice Conversion in Codec Space}, 
      author={Qixi Zheng and others},
      year={2026},
      journal={arXiv preprint arXiv:2604.12456},
}

@article{song2025stabletoken,
      title={StableToken: A Noise-Robust Semantic Speech Tokenizer for Resilient SpeechLLMs}, 
      author={Yuhan Song and others},
      year={2026},
      journal={arXiv preprint arXiv:2509.22220},
}

@article{chen2025sac,
      title={SAC: Neural Speech Codec with Semantic-Acoustic Dual-Stream Quantization}, 
      author={Wenxi Chen and others},
      year={2025},
      journal={arXiv preprint arXiv:2510.16841},
}

@article{tao2024tone,
      title={ToneUnit: A Speech Discretization Approach for Tonal Language Speech Synthesis}, 
      author={Dehua Tao and others},
      year={2024},
      journal={arXiv preprint arXiv:2406.08989}
}

\end{document}